\documentstyle[eqsecnum,aps,preprint]{revtex}
\begin{document}
\draft
\title{Towards a fully self-consistent spectral function of the nucleon
in nuclear matter}

\draft

\author{F. de Jong and H. Lenske}
\address{Institut f\"ur Theoretische Physik, Universit\"at Giessen,
35392 Giessen, Germany}
\date{\today}

\maketitle
\begin{abstract}
We present a calculation of nuclear matter which goes beyond the
usual quasi-particle approximation in that it includes part of the
off-shell dependence of the self-energy in the self-consistent solution
of the single-particle spectrum.
The spectral function is separated in contributions for energies above and
below the chemical potential.
For holes we approximate the spectral function for energies below 
the chemical potential by a $\delta$-function at the quasi-particle peak 
and retain the standard form for energies above the chemical
potential. 
For particles a similar procedure is followed.
The approximated spectral function is consistently used at all levels 
of the calculation. 
Results for a model calculation are presented, the main conclusion is
that although several observables are affected by the 
inclusion of the continuum contributions the physical consistency of the model 
does not improve with the improved self-consistency of the solution method. 
This in contrast to expectations based on the crucial role of 
self-consistency in the proofs of conservation laws.
\end{abstract}


\section{Introduction}
Nuclear matter provides an ideal testing ground for the theory of 
strongly interacting many-body systems. 
Translational invariance leads to a particulary simple ground state
structure where wave functions are given by plane waves. 
This allows to focus on the specific many-body scheme. 
Brueckner already identified the leading contributions by noting
that the dynamics are dominated by contributions from ladder-type 
diagrams. 
Further progress was achieved by embedding of the equilibrium description
of the ground state in a dynamic non-equilibrium theory \cite{Baym}.
This introduces the conservation laws of e.g. energy, particle number and
momentum, even in a theory which only deals with static equilibrium 
where no explicit time development is present.
Kadanoff and Baym showed that the requirement that the 
approximated theory preserves the conservation laws is not trivial \cite{Baym}.
Closer examination reveals that the Brueckner approach is a subset of 
a 'conserving' approximation, the so-called $T$-matrix approximation
\cite{KoehlerII}.
The first term of the Brueckner expansion ignores hole-hole 
propagation, which turn out to be crucial in preserving the
conservation laws. 
Including these additional terms, as done e.g. in a 
non-relativistic approach in Ref. \cite{Ramos} and in a relativistic 
model in Refs. \cite{FdJ_cons}, indeed shows a marked improvement in the
physical consistency of the description of the many-body system 
and the preservation of conservation laws.
However, deviations still persist and it is both necessary and interesting
to investigate the causes.

A crucial aspect in the proofs that the approximated theory preserves
the conservation laws is the self-consistency of the solution of the system.
To make the solution of the $T$-matrix approximation feasible one has
to resort to several additional practical approximations.
Note the essential difference between the approximation in the many-body
scheme, as e.g. introduced by employing the $T$-matrix approximation,
in contrast to the practical approximations needed to make a calculation
feasible.
One common simplification is to approximate the intermediate
2-body propagators by an angle-averaged form \cite{Brueckner_avg}. 
Then one can decompose the T-matrix in partial waves and the spatial 
integration reduces to a one-dimensional integration over the momentum
variable and the full T-matrix is reconstructed by a sum over the 
separate partial wave contributions.
As shown in Refs. \cite{Cheon,Sartor} this is a reliable approximation.

The integrations over the energy variable are usually simplified by 
assuming that the imaginary part of the self-energy is zero and taking 
for the real part the value of the self-energy at the pole of the propagator. 
The spectral function then reduces to a $\delta$-function with its peak at
the quasi-particle pole.
This ensures that in this 'quasi-particle' approximation
all energy integrations are easily carried out analytically.
Together with the angle-averaging procedure the problem is then reduced
to a self-consistent solution of the self-energy which now only depends
on the magnitude of the momentum.

In the self-consistent solution in the quasi-particle approximation of 
the $T$-matrix only the values 
of the self-energies at the quasi-particle peak are calculated.
Clearly, a calculation of the full energy-momentum dependence of the 
self-energy will check the validity of the quasi-particle approximation.
If one finds a small imaginary part and therefore
a weak dependence of the real part on the energy variable
the quasi-particle approximation is probably justified.  
This was done at zero temperature in a non-relativistic approach by e.g. 
Ramos et al. \cite{Ramos}, in  Ref. \cite{FdJ_cons,FdJ_consII} results
for a relativistic Dirac-Brueckner model were presented.
The Rostock group performed non-relativistic calculations at finite 
temperature, focussing on bound states and critical temperatures \cite{Alm}.
All these calculations essentially find that most of the strength of the 
spectral function is indeed centered around the quasi-particle peak. 
However, a significant percentage (for nuclear matter in the range of 10-20\%, 
depending on the model) is moved to energies farther out.
Since the equations are non-linear it is not a priori clear what the effects
of incorporating this in the solution would be. 
In the present work we present an approximation of the 
spectral function that does include part of the off-shell dependence
of the self-energy.
We will investigate the effect of it on various observables in a 
practical calculation for a model interaction and compare the results
with those of the quasi-particle approximation. 
The organisation of this paper is as follows. 
The next section is devoted to the theoretical framework, in the
third section we present results for a calculation using a model 
potential, while in the final section we will summarize our results
and draw conclusions.

\section{Theoretical Framework}

A complete account of the employed formalism and the philosophy behind 
it can be found in Ref. \cite{Botermans}.
The central quantity in the many-body theory is the self-energy.
It describes the effects of the medium and
contains all the information on 1-body properties of the system.
In uniform equilibrated nuclear matter it is a function of both the 
momentum and the energy. 
As any physically relevant function, it is an analytic function of
the energy. 
Therefore the real and imaginary part of the self-energy are 
related by a dispersion relation. 
Since it is impossible to calculate the full theory, one has to 
resort to an approximation which sums a specific class of diagrams.
In Refs. \cite{Baym,Botermans} a method of obtaining systematic  
approximations for the self-energy is presented. 
One first establishes a Martin-Schwinger hierarchy of n-body Green's
functions.
These express lower-n Green's functions in higher-n ones.
One then defines a formal self-energy in which all the information on
correlations is present.
The self-energy can be approximated in several ways, 
thereby truncating the Martin-Schwinger hierarchy and making the system solvable.
Clearly, for the approximation of the self-energy 
constraints are set by the physical characteristics of the system. 
Several schemes that also respect conservation laws ('conserving approximations')
have been discussed in the literature \cite{Baym,Botermans}.
For nuclear matter with its strong short-range interaction the $T$-matrix
approximation is appropriate.
This is closely related to the Brueckner approach of summing ladder-type diagrams, but 
the T-matrix approximation goes beyond this by including also the 
propagation of hole-pairs. 
This is essential in fulfilling conservation laws \cite{Baym,Botermans}.

We now discuss to the representation of the $T$-matrix approximation in
momentum space.
We first introduce the propagators, using the Dyson equation we obtain  
the effective (dressed) advanced propagator in the well known form:
\begin{equation}
g^{(\pm)}(\omega, \bar{p}) = \frac{1}{\omega - \frac{p^2}{2m} - \Sigma^{(\pm)}(\omega, \bar{p})},
\label{def_gplus}
\end{equation}
where the retarded self-energy is defined as 
$\Sigma^{(+)}(\omega, \bar{p}) = \Sigma(\omega - i \eta, \bar{p}), \eta \rightarrow 0^+$.
Inserting this in the definition of the spectral function one obtains the standard form
\begin{eqnarray}
a(\omega, \bar{p}) &=& g^{(+)}(\omega, \bar{p}) - g^{(-)}(\omega, \bar{p}) \nonumber\\
&=& 
\frac{-2 Im \left[\Sigma^{(+)}(\omega, \bar{p}) \right]}
{(\omega - \frac{p^2}{2m} - Re \left[\Sigma^{(+)}(\omega, \bar{p}) \right])^2 + 
 (Im \left[\Sigma^{(+)}(\omega, \bar{p}) \right])^2}.
\end{eqnarray}
At temperature $T$ the correlation propagators can be expressed in terms of 
the spectral function and the Fermi distribution 
$f(\omega) = (\exp((\omega - \mu)/T + 1)^{-1}$ where $\mu$ is the chemical potential;
\begin{eqnarray}
g^<(\omega, \bar{p}) &=&  f(\omega) a(\omega, \bar{p}) \nonumber\\
g^>(\omega, \bar{p}) &=&  (1 - f(\omega)) a(\omega, \bar{p}).
\label{def_gcorr}
\end{eqnarray}
The set of propagators Eqs. \ref{def_gplus},\ref{def_gcorr} 
fully describes the 1-particle properties of the 
system.
In equilibrium they are fully determined by the retarted self-energy 
$\Sigma^{(+)}(\omega, \bar{p})$, reflecting again that the self-energy 
contains all the physical information of the system.
In other words, in equilibrated nuclear matter only one independent function 
describes the system: specifying the spectral function determines the 
self-energy and vice-versa.

In the $T$-matrix approximation the self-energy is given by (we use a notation where 
$p$ is the four-momentum specified by the pair $(\omega, \bar{p})$ and $\bar{p}$ denotes
the usual 3-momentum
\begin{eqnarray}
\Sigma^{(+)}(\omega, \bar{p}) &=& \int \frac{d^4p'}{(2 \pi)^4} 
\Big[
\langle \frac{1}{2}(\bar{p} - \bar{p}') | T^{(+)}(\Omega, \bar{P}) |  
\frac{1}{2}(\bar{p} - \bar{p}') \rangle_{A}
g^<(\omega', \bar{p}') \nonumber \\
& & + \langle \frac{1}{2}(\bar{p} - \bar{p}') | T^<(\Omega, \bar{P}) |  
\frac{1}{2}(\bar{p} - \bar{p}') \rangle_{A} g^{(-)}(\omega', \bar{p}') 
\Big],
\label{def_sigma}
\end{eqnarray}
where $\Omega = \omega + \omega', \bar{P} = \bar{p} + \bar{p}'$.
The T-matrices are given by:
\begin{equation}
\langle \bar{p} | T^{(+)} (\Omega, \bar{P}) | \bar{p}' \rangle =
\langle \bar{p} | V | \bar{p}' \rangle -
\int \frac{d^3p''}{(2\pi)^3} 
\langle \bar{p} | V | \bar{p}'' \rangle 
g_{12}^{(+)}(\Omega, \bar{P}, \bar{p}'')
\langle \bar{p}'' | T^{(+)} (\Omega, \bar{P}) | \bar{p}' \rangle,
\label{def_tmat}
\end{equation}
with the propagator
\begin{equation}
g_{12}^{(+)}(\Omega, \bar{P}, \bar{p}) =
\int \frac{d\omega_1 d\omega_2}{(2\pi)^2} 
\frac{g^>(\omega_1, \frac{\bar{P}}{2} + \bar{p})
      g^>(\omega_2, \frac{\bar{P}}{2} - \bar{p}) -
      g^<(\omega_1, \frac{\bar{P}}{2} + \bar{p})
      g^<(\omega_2, \frac{\bar{P}}{2} - \bar{p})}
{\Omega - \omega_1 - \omega_2 + i \epsilon}
\end{equation}
and the correlation $T$ matrix
\begin{equation}
\langle \bar{p} | T^< (\Omega, \bar{P}) | \bar{p}' \rangle =
\int  \frac{d^4p''}{(2\pi)^4}
\langle \bar{p} | T^{(+)} (\Omega, \bar{P}) | \bar{p}'' \rangle
g^<(\omega'', \bar{p}'') g^<(\Omega - \omega'', \bar{P} - \bar{p}'')
\langle \bar{p}'' | T^{(+)} (\Omega, \bar{P}) | \bar{p}' \rangle.
\end{equation}
The set of coupled equations for the self-energy and the $T$-matrix 
are too complicated to solve in its full dependence on the spatial 
momentum and energy variable and one has to resort to
the conventional approximations of angle-averaging
the two-body propagator and the quasi-particle approximation 
discussed before.

The equations above display the extensions of the $T$-matrix approximation
over the Brueckner scheme in that it also propagates hole-pairs. 
This is also expressed in the equations above.
The Brueckner scheme is an expansion in terms of the density, in our
formulation this is the term in first order of $g^<$.
Brueckner theory is retrieved by neglecting all higher order terms in 
$g^<$ to zero: 
the hole-hole propagating parts in  $g_{12}$ (the terms with products
of $g^<$, they are of second order in $g^<$) and $T^<$.
Then $g_{12}$ only propagates particle-particle terms and
$T^{(+)}$ only contains particle-particle ladder diagrams.
Only the first term in the definition of Eq. \ref{def_sigma} contributes
to the self-energy which is equal to the conventional Brueckner self-energy.
A detailed account of the relation between the Green's function
approach used in this work and Brueckner theory is found in 
Ref. \cite{KoehlerII}.

\subsection{Sum-rule conserving approximations of the spectral function}

In the present study we go beyond the quasi-particle approximation
and by including part of the off-shell energy dependence of the
self-energy.
In approximating the spectral function one has to be careful not to
introduce unphysical aspects. 
One important property of the spectral function is the sum-rule:
\begin{equation}
\int \frac{d\omega}{2\pi} a(\omega, \bar{p}) = 1.
\label{def_sumrule}
\end{equation}
This just states that the spectral function is a probability 
distribution normalized to one. 
The interaction spreads the probability of finding a 
particle with a given momentum over the full energy range. 
The sum-rule ensures that the total strength is preserved. 
It is a straightforward consequence of the anti-commutation relations
of the fields. 
However, it is also easy to prove that a sufficient condition for Eq. 
\ref{def_sumrule} is that the underlying self-energy is 
analytic. 
If the imaginary part of the retarded self-energy is negative for all
$\omega$ in the upper half-plane, all poles of the retarded propagator
will lie in the lower half-plane. From the definition of the spectral
function it is clear that the sum of all the residues is then equal
to the sum-rule. 
Using standard complex analysis this is equal to the residue at infinity,
which is easily proven to be unity for physically reasonable 
self-energies. 
Hence, any physically meaningful approximation of the spectral function must
conserve the sum-rule. 
If the sum-rule is violated (as e.g. in Ref. \cite{Weigel}), 
the underlying self-energy is not analytic and thus not physical.
Another good illustration of this point is the so-called Extended Quasi Particle
approximation \cite{Kraeft,Koehler}. 
Essential in the proof that this approximation fulfills the sum-rule 
is the analyticity of the self-energy. 

Stability of the ground state at $T = 0$ requires that the imaginary part of
the self-energy at $\omega = \mu$ vanishes
\begin{equation}
Im[ \Sigma^{(+)}(\mu, \bar{p})] = 0.
\label{zero_im_sig}
\end{equation}
This results in a zero for the spectral function at $\omega = \mu$.
As was shown in Ref. \cite{Alm} this property disappears for finite 
temperature, although remnants can still be seen at quite high energies. 
As we will see later, due to the analyticity of the self-energy
this also has repercussions on the real part of the self-energy.
At $p = p_f$ it results in the Fermi-surface consisting of 
ideal quasi-particles with infinite lifetime. 
This reflects the fact that the calculated groundstate has 
to be stable:
although all states in the Fermi sea have a finite lifetime, their
decay is Pauli-blocked. 
This is directly related to the finding that the collision term in 
the equation of motion is zero in equilibrated matter at $T=0$ \cite{Botermans}. 
Only the decay of particles at the Fermi surface is not Pauli blocked,
but they have an infinite lifetime and the system is thus stable.

The Extended Quasi Particle approximation (EQP) is an attractive candidate 
for a practical approximation of the spectral function that includes
off-shell effects. 
Almost all of the energy integrations can be carried out analytically. 
Also the zero of the spectral function at $\omega = \mu$ is preserved
which is the case for more schematic approximations as e.g. using 
a single Lorentzian with the real and imaginary parts of the
self-energy fixed to the values at the quasi-particle peak.
However, as we will see later in a calculation, for a strong enough
interaction the quasi-particle
strength becomes larger than unity for low momenta. 
This is a consequence of the imaginary part of the self-energy 
being zero at $\omega = \mu$ and the self-energy being analytic.
This feature makes the EQP difficult to implement in a nuclear medium.

We seek an approximation that introduces off-shell effects as 
a perturbation on the quasi-particle solution. 
This is achieved by approximating the spectral function
separately for particle and holes. 
We define particles as having momenta for which the quasi-particle peak 
lies above $\mu$ and holes have a quasi-particle peak below $\mu$.
In practice this amounts to splitting the spectral function 
into distinct parts for momenta higher or lower than the Fermi-momentum,
respectively.
We define a 'quasi-hole' by assinging the full hole-strength with energies
smaller then $\mu$ into a single quasi-particle peak and retain the 
standard spectral function for energies above $\mu$.
Since the spectral function is zero at $\omega = \mu$ the 
approximated spectral function is still a continuous function of the energy.
For particles we apply a similar procedure and we get
\begin{eqnarray}
a_h(\omega, \bar{p}) &=& 
Z_h 2\pi \delta(\omega - \omega_{qp}) + \theta (\omega - \mu) a(\omega, \bar{p})
\nonumber\\
a_p(\omega, \bar{p}) &=& 
\theta (\mu - \omega) a(\omega, \bar{p}) + Z_p 2\pi \delta(\omega - \omega_{qp})
\nonumber\\
\tilde a(\omega, \bar{p}) &=& \theta(p_f - p) a_h(\omega, \bar{p}) + 
\theta(p - p_f) a_p(\omega, \bar{p}).
\label{def_app_spec}
\end{eqnarray}
The factors $Z_{p,h}$ are determined by requiring that the approximated spectral function 
fulfills the sum-rule for all momenta. 
In the equations $\omega_{qp}$ denotes the 
position of the quasi-particle peak: the energy at which the real part of the
denominator of the propagator vanishes.

By construction the approximated spectral function conserves the same numerical values for 
the momentum distribution $n(\bar{p})$, defined by
\begin{equation}
n(\bar{p}) = \int_{-\infty}^\mu d\omega a(\omega, \bar{p}),
\label{def_momentum}
\end{equation}
and the factors $Z_{p,h}$ are simply expressed in terms of $n(p)$:
$Z_h = n(\bar{p}), Z_p = 1 - n(\bar{p})$.

In the system and with the model interaction we consider in this work
a few percent of the total strength is moved across $\mu$, so the 
$Z$-factors are close to one and the quasi-particle peak still carries the 
bulk of the strength. 
This is exactly what we intended, namely to introduce the off-shell effects as 
a perturbation on the standard calculation in the quasi-particle approximation.

\subsection{Calculation of propagators and the T-matrices}

Inserting the approximated spectral function of Eq. \ref{def_app_spec} 
in the correlation propagators we obtain
\begin{eqnarray} 
g^<(\omega, \bar{p}) &=& \theta(\mu - \omega) 
\left[
\theta(p_f - p) Z_h(p) 2\pi \delta(\omega - \omega_{qp}) + 
\theta(p - p_f) a(\omega, \bar{p})
\right]
\nonumber \\
g^>(\omega, \bar{p}) &=& -\theta(\omega - \mu) 
\left[
\theta(p - p_f) Z_p(p) 2\pi \delta(\omega - \omega_{qp}) + 
\theta(p_f - p) a(\omega, \bar{p})
\right]
\end{eqnarray}
With these forms we can work out the two-body propagator in Eq. \ref{def_tmat}.
We find contributions of products of $\delta$-functions, products of a 
$\delta$-function with a tail of a spectral function and products of tails.
For numerical simplicity we drop the latter contributions since they carry
only a very small portion of the total strength
(they are of order $O((1 - n(p))^2)$ for holes).
We have
\begin{eqnarray}
g_{12}^{(+)}(\Omega, \bar{P}, \bar{p}) &=&
\frac{Z_h(p_1) Z_h(p_2) \theta(p_f - p_1) \theta(p_f - p_2) - 
      Z_p(p_1) Z_p(p_2) \theta(p_1 - p_f) \theta(p_2 - p_f)}
{\Omega - \omega^1_{qp} - \omega^2_{qp} + i \epsilon}
\nonumber \\
&+& \int_{-\infty}^{\mu} \frac{d \omega_1}{2\pi}
\frac{a(\omega_1, \bar{p}_1) Z_h(p_2) \theta(p_1 - p_f) \theta(p_f - p_2)}
{\Omega - \omega_1 - \omega^2_{qp} + i \epsilon}
\nonumber \\
&+& \int_{-\infty}^{\mu} \frac{d \omega_2}{2\pi}
\frac{Z_h(p_1) a(\omega_2, \bar{p}_2) \theta(p_f - p_1) \theta(p_2 - p_f)}
{\Omega - \omega^1_{qp} - \omega_2 + i \epsilon}
\nonumber \\
&-& \int^{\infty}_{\mu} \frac{d \omega_1}{2\pi}
\frac{a(\omega_1, \bar{p}_1) Z_p(p_2) \theta(p_f - p_1) \theta(p_2 - p_f)}
{\Omega - \omega_1 - \omega^2_{qp} + i \epsilon}
\nonumber \\
&-& \int^{\infty}_{\mu} \frac{d \omega_2}{2\pi}
\frac{Z_p(p_1) a(\omega_2, \bar{p}_2) \theta(p_1 - p_f) \theta(p_f - p_2)}
{\Omega - \omega^1_{qp} - \omega_2 + i \epsilon},
\label{g12_app}
\end{eqnarray}
where we used the shorthand notation $\bar{p}_{1,2} = \frac{\bar{P}}{2} \pm \bar{p}$.
On this propagator we apply a standard angle-averaging procedure
\cite{Brueckner_avg}.
All factors including $\theta$-functions in the momenta will reduce to the
standard angle-averaged Pauli operators \cite{Ramos}, $Q_{pp}, Q_{hh}$ and $Q_{ph} = 
1 - Q_{pp} - Q_{hh}$.
In a similar fashion also the Z-factors and the self-energies 
are replaced by angle-averaged values.
The propagator then becomes
\begin{eqnarray}
g_{12}^{(+)}(\Omega, \bar{P}, \bar{p}) &=&
\frac{Z_{hh}(P,p) Q_{hh}(P, p) - Z_{pp}(P,p)Q_{pp}(P, p)}
{\Omega - \frac{P^2}{4m} - \frac{p^2}{m} - 2\Sigma(P,p)_{qp} + i \epsilon}
\nonumber \\
&+& \int_{-\infty}^{\mu} \frac{d \omega_2}{\pi}
\frac{Z_h(p_1) a(\omega_2, \bar{p}_2) Q_{ph}(P,p)}
{\Omega - \omega^1_{qp} - \omega_2 + i \epsilon}
\nonumber \\
&-& \int^{\infty}_{\mu} \frac{d \omega_2}{\pi}
\frac{Z_p(p_1) a(\omega_2, \bar{p}_2) Q_{ph}(P,p)}
{\Omega - \omega^1_{qp} - \omega_2 + i \epsilon}.
\label{g12_app_ang}
\end{eqnarray}
We will refer to the first part as the quasi-particle propagator,
to the contributions from the integrations as the continuum 
contributions.
The continuum contributions vanish for $Im (\Sigma) = 0$,
in that case all $Z_{p,h}$ factors are equal to unity
and the quasi-particle propagator
reduces to the conventional angle-average two-body propagator as
e.g. used in Refs. \cite{Ramos,Alm}.
Angle-averaging the continuum contributions is achieved in the following way:
for a given pair of $(P,p)$ where $Q_{ph}(P,p)$ is non-zero
the pair $(p_1,p_2)$ is formed for an intermediate value of
the angle such that $p_1 < p_f < p_2$. 
Then the integration over the energy variable is performed. 
Note that by construction the tails of the spectral functions
appearing in the integrals are smooth functions. 

The imaginary part of the two-body propagator has now two contributions,
one from the elastic cut in the quasi-particle propagator, 
the other from the imaginary parts of the continuum
contributions. This is reflected by the unitarity relation
obeyed by the T-matrix.
From the T-matrix equation one derives the formal
unitarity relation in partial-wave representation
(in the case of coupled channels the summation over these
in the intermediate states is implicit):
\begin{equation}
Im[ \langle p | T^{(+)}(\Omega, P) | p' \rangle] =
\int \frac{dp'' {p''}^2}{\pi}
\langle p | T^{(+)}(\Omega, P) | p'' \rangle
Im[g_{12}(\Omega, P, p'')]
\langle p'' | T^{(-)}(\Omega, P) | p' \rangle.
\end{equation}
Using Eq. \ref{g12_app_ang} this relation can easily be
worked out. 
The contribution from the quasi-particle part is trivial.
Taking the imaginary part of the continuum contributions 
the denominator under the integral reduces to a $\delta$ function 
which sets $\omega_2 = \Omega - \omega^1_{qp}$
and we obtain
\begin{eqnarray}
\lefteqn{Im[ \langle p | T^{(+)}(\Omega, P) | p' \rangle] =} 
\nonumber\\
&-& \frac{p_0^2 (Z_{hh}(P,p_0) Q_{hh}(P, p_0) - Z_{pp}(P,p_0)Q_{pp}(P, p_0))}
{2 \frac{p_0}{m} + 2\Sigma'(P,p_0)_{qp}}
\langle p | T^{(+)}(\Omega, P) | p_0 \rangle
\langle p_0 | T^{(-)}(\Omega, P) | p' \rangle
\nonumber\\
&-& \int \frac{dp'' {p''}^2}{\pi}
Q_{ph}(P,p'') a(\Omega - \omega^1_{qp}, p_2)
\left[
Z_h(P,p_1) \theta(\mu - \Omega +  \omega^1_{qp}) -
Z_p(P,p_1) \theta(\Omega -  \omega^1_{qp} - \mu) 
\right]
\nonumber\\
&\times&
\langle p | T^{(+)}(\Omega, P) | p'' \rangle
\langle p'' | T^{(-)}(\Omega, P) | p' \rangle,
\label{unitarity}
\end{eqnarray}
where $p = p_0$ denotes the momentum at the quasi-particle peak of
$g_{12}$:
$\Omega + \frac{P^2}{4m} + \frac{p^2_0}{m} + 2\Sigma(P,p_0)_{qp} = 0$.
$\Sigma'(P,p_0)_{qp}$ is the derivative with respect to the 
momentum of the self-energy:
\begin{equation}
\Sigma'(P,p_0)_{qp} = \frac{\partial}{\partial p} \Sigma(\omega_{qp},P,p) |_{p_0}.
\end{equation}
At temperature $T=0$ and $\Omega = 2 \mu$, $p_0$ has a value such that 
$Q_{hh}(P, p_0) = Q_{pp}(P, p_0) = 0$ and the 
quasi-particle part does not generate an imaginary part.
The same is true for the continuum part at $\Omega = 2 \mu$.
In the first $\theta$-function, $\omega^1_{qp} < \mu$ and the
$\theta$-function is zero. 
A similar argument holds for the second $\theta$-function.
Altogether, this results in $Im[T(\Omega = 2 \mu)] = 0$. 
This leaves the possibility that the system develops 
a two-body bound state at this energy, which can only happen
for $Im(T) = 0$.
This was e.g. explored in Ref. \cite{Alm}. 

At $T = 0$ the unitarity relation, Eq. \ref{unitarity}, 
provides a simple relation between $T^<$ and $T^{(\pm)}$ 
\begin{equation}
\langle p | T^<(\Omega, P) | p \rangle = -
\theta(2 \mu - \Omega) Im [ \langle p | T^{(+)}(\Omega, P) | p \rangle ],
\end{equation}
and using Eq. \ref{g12_app_ang} one finds
\begin{eqnarray}
\lefteqn{\langle p | T^<(\Omega, P) | p \rangle =}
\nonumber \\
&\phantom{+}& \frac{p_0^2 Z_{hh}(P,p_0) Q_{hh}(P, p_0)}
{2 \frac{p_0}{m} + 2\Sigma(P,p_0)'_{qp}}
\langle p | T^{(+)}(\Omega, P) | p_0 \rangle
\langle p_0 | T^{(-)}(\Omega, P) | p' \rangle
\nonumber\\
&+& \int \frac{dp'' {p''}^2}{\pi}
Q_{ph}(P,p'') a(\Omega - \omega^1_{qp}, p_2)
Z_h(P,p_1) \theta(\mu - \Omega +  \omega^1_{qp})
\nonumber\\
&\times&
\langle p | T^{(+)}(\Omega, P) | p'' \rangle
\langle p'' | T^{(-)}(\Omega, P) | p' \rangle.
\label{t_small}
\end{eqnarray}
Again, we refer to the first part as the quasi-particle contribution and
to the second as the continuum contribution.

\subsection{Evaluation of self-energies}

In calculating the self-energy we of course have to use the same
approximated spectral function consistently in all expressions. 
In the first term of Eq. \ref{def_sigma}, $T^{(+)}g^<$,
this is straightforward.
For the second term the fact that 
$Im[ T^< g^{(-)}(\omega, \bar{p})] = \frac{1}{2} T^< a(\omega, \bar{p})$
allows us to use the approximated spectral function. 
This provides us with the imaginary part of this contribution.
Since both contributions are analytic functions we can then use a
dispersion relation to calculate the real part of 
$T^< g^{(-)}(\omega, \bar{p})$. 
This is facilitated by the fact that the imaginary part of this
contribution is non-zero over a range of only a few hundred 
MeV around the Fermi energy.
We can identify several contributions to the self-energy, originating
from various combinations of continuum and quasi-particle parts. 
Explicit expressions and more details of the calculation of
the self-energy are given in the appendix.

Now, we are almost in a position to calculate the self-energy. 
In the expressions given below we need $T^{(+)}$ at all possible
values of the variables $(\Omega, P, p)$. 
This we achieve by a suitable interpolation. 
We separate the interpolation in one for the real part of $T^{(+)}$, and, 
using the unitarity relation the interpolation of the imaginary part
is split into a quasi-particle part and a continuum contribution. 
In the interpolation of the quasi-particle part the
factor with the Pauli-blockers in front is separated out.   
These trivial factors, which are easily calculated, depend strongly on the 
momenta and hinder a good interpolation numerically. 
It is more efficient to interpolate the term $T^{(+)}T^{(-)}$ and 
multiply with the factors later on, which greatly improves the 
stability of the results. 

After obtaining a self-consistent solution of the self-energy we
can calculate the binding energy and other observables. 
The energy of the system is given by 
\begin{eqnarray}
E &=& I\frac{1}{2} \int \frac{d^4p}{(2\pi)^4} 
(\frac{\bar{p}^2}{2m} + \omega) g^<(\omega, \bar{p})
\nonumber \\
&=& 
I\int_0^{\infty} \frac{d^3p}{2(2\pi)^3} \frac{\bar{p}^2}{2m} n(\bar{p})
+ I\int_0^{p_f} \frac{d^3p}{2(2\pi)^3} 
\left(
\frac{\bar{p}^2}{2m} + \Sigma^{(+)}(\omega_{p,qp},\bar{p})
\right)
n(\bar{p}) 
\nonumber \\
&+& I\int_{p_f}^{\infty} \frac{d^3p}{2(2\pi)^3}
\int_{-\infty}^{\mu} \frac{d\omega}{2\pi} \, \omega \, a_p(\omega,\bar{p}),
\label{def_energy}
\end{eqnarray}
where $I$ is the spin-isospin degeneracy factor,
$I=2$ for neutron matter and $I=4$ for nuclear matter, respectively.
The particle-density of the system is given by:
\begin{equation}
\rho = I \int \frac{d^4p}{(2\pi)^4} g^<(\omega, \bar{p}) =
I \int \frac{d^3p}{2(2\pi)^3} n(\bar{p}).
\label{def_density}
\end{equation}
In the quasi-particle approximation (and therefore also for a 
non-interacting system) this gives the standard relation between 
the Fermi momentum and the density: $\rho = \frac{I}{6\pi^2} p^3_f$.
A conserving approximation conserves the total particle number, and 
after switching on the interaction one must obtain the same result for 
the density. 
Deviations from this ideal situation can have various causes. 
Excluding trivial numerical inaccuracies, deviations indicate 
insufficiencies of the approximation scheme. 
We stressed before that a crucial requirement for a conserving approximation
is self-consistency. 
Although we improved upon this by including off-shell effects,
our approach is still not fully self-consistent since we still ignore
the width of the quasi-particle peak in our approximated spectral
function.
This might introduce deviations from the ideal particle 
conservation.
On the other hand, in momentum space the conservation laws are only
proven in the gradient expansion, which ignores gradients
of the $T$-matrix in the Wigner transform \cite{Botermans}. 
So violations of the conservation laws also might point
to inaccuracies introduced by the gradient expansion. 
Note however that we still can define an unique value of the Fermi-momentum
by setting it to the momentum for which the quasi-particle energy is 
equal to the chemical potential.

\section{Results}

\subsection{Details of the numerical calculation}

All the results presented in this study were calculated with a non-relativistic
version of the Bonn-C potential \cite{Machleidt}, some numerical details are 
discussed in the appendix.
This provides an adequate fit to the phase-shifts and leaves the 
possibility of exploring relativistic effects in the 
future. 
To keep the numerical effort within limits we have to restrict ourselves
to the most important partial-wave contributions. 
In the isospin $T = 1$ channel we keep the $^1S_0$ partial wave, in the
isospin $T = 0$ channel the coupled-channel ${^3S_1}-{^3D_1}$ partial wave.
The tensor force of the latter channel has two effects. 
Firstly, it is very effective in moving strength of the spectral 
function across $\omega = \mu$, thereby reducing $n(\bar{p})$ and
increasing the importance of the tail of the spectral function.
On the other hand, however, it is also responsible for the presence
of a bound state, discussed extensively in Refs. \cite{Ramos,Alm,VonderfechtI,VonderfechtII}.
Although a treatment of this 'deuteron-like' bound state is very well 
possible within our approximation scheme, we leave this for a future study.
Since we are interested in the effects of the continuum contributions, we
want to keep as much of the tensor force as possible.
Reducing its strength by a factor 1/2 removes the bound state 
%
%
and we can perform a calculation in
a range of densities around the nuclear matter saturation density.

We first calculate a self-consistent spectrum within 
the quasi-particle approximation. 
Then, still in the quasi-particle approximation,
the self-energies are calculated on a mesh in the energy-momentum plane,
which, in turn, is used as a starting point for the full calculation
including the continuum contributions. 
For nuclear matter it turns out that for an adequate fulfillment of the
sum-rule Eq. \ref{def_sumrule} we need to include energies up to 
about 2 GeV.
Using the change in the resulting quasi-particle spectrum as an 
iteration parameter we find rapid convergence:
for depletions in the order of 6 \% two iterations are usually 
sufficient.

\subsection{Quasi-particle spectrum and self-energies}

The resulting quasi-particle spectra for a calculation at 
normal nuclear matter density ($p_f = 0.27$ GeV/c) are displayed in 
Fig. \ref{qp_spectrum}.
Deep in the Fermi sea the real part of the self-energy in the 
continuum calculation is some 3 MeV less negative than the one 
calculated in the quasi-particle approximation.
For larger momenta the difference gets smaller: at the Fermi
momentum the difference is around 2 MeV, at twice the Fermi
momentum it is negligible.
For the imaginary parts we again find the largest difference 
for small momenta.
The imaginary part of the self-energy in the continuum calculation
is 1.5 MeV smaller (less negative) than in the quasi-particle 
approximation.
Approaching the Fermi momentum the difference obviously 
vanishes since the quasi-particle energy at the Fermi surface
equals the chemical potential resulting in a vanishing  
imaginary part of the self-energy. 
For higher momenta we do not find significant differences
within the calculated range. 

In figure \ref{full_sig} we show the full dependence on both 
the energy variable $\omega$ and the momentum $\bar{p}$ of the 
real and imaginary parts
of the self-consistent self-energy calculated at normal nuclear
matter density. 
For a better presentation the projection angles are taken differently
and the imaginary part is multiplied by a factor -1.
At higher energies the imaginary part shows a similar behaviour 
for all momenta. 
The zero of the imaginary part at $\omega = \mu$ is clearly visible.
At lower momenta we observe a peak, 
this is due to the large phase shifts for low laboratory
energies in the $L = 0$ partial waves. 
The imaginary part can be viewed as a cross section averaged over
the momenta. 
In the equation of motion these are multiplied by phase-space
factors which ensure Pauli-blocking.
For low $p_{lab}$ one predominantly averages over low total momenta, 
these are equivalent to low laboratory energies and one picks
up the large cross sections. 
At higher $p_{lab}$ the average is over higher momenta where
the cross-sections are smaller. 
Together with the requirement that 
$Im[\Sigma(\omega = \mu)] = 0$
one obtains the structure as observed in the figure. 
Since the self-energy is an analytic function and therefore 
observes a dispersion relation the structure seen in the 
imaginary part is reflected in the real part.
This is clearly the case, with the largest structure being
present at the lower momenta. 
For momenta around the Fermi momentum the quasi-particle point 
lies at the left side of the ridge (from the point of view 
in the figure), for lower momenta it lies on the right side
of the ridge.
In the latter case, the derivative of the real part of the
self-energy with respect to the energy is positive, leading to a 
quasi-particle strength larger than one. 
This is reflected by the spectral function having two maxima, 
centered around the quasi-particle pole. 
If the derivative gets larger then unity the spectral function 
even has two quasi-particle poles. 
In our system this can happen at lower densities, there the average of 
the momentum is taken over a smaller volume, leading to a larger 
imaginary part.
Also the energy scale is compressed a bit (i.e. the maximum in the
imaginary part lies closer to $\omega = \mu$). This all
leads to a larger positive derivative at low momenta. 
As we already noted before this makes the application of the 
Extended Quasi-Particle approximation in a scheme similar to 
the one we employ problematic.
A similar behaviour was found in Refs. \cite{Alm,Koehler}, and
with the arguments given above
we think this is an unavoidable feature of any realistic interaction 
having large cross-sections at low energies. 

In Fig. \ref{sig_omega} we compare the energy dependence of the
self-energy at a fixed momentum. 
We now use the variable $\omega' = \omega - \omega_{qp}$, defined
as the distance to the quasi-particle energy. 
This allows for a meaningful comparison between the results, even when
the underlying self-consistent quasi-particle spectra are different.
The imaginary parts of the self-energies calculated in both approaches
show only minor differences on the scale of the figure. 
Only the maximum below $\omega' = 0$ is slightly larger for the
quasi-particle calculation. 
The difference between the real parts is nearly constant (as one would
expect when the imaginary parts do not differ on the same scale).
This is caused by a smaller imaginary part of the self-energy in the
continuum calculation for higher energies. 

\subsection{Binding energy and other observables}

Having found a self-consistent self-energy we can calculate
the whole set of one-particle observables.
In Tables \ref{qp_observables} and \ref{cont_observables} we present
values for several observables calculated in the quasi-particle and
in the continuum method of this paper, respectively.
In the second column the binding energy is shown.
For the continuum method these are in the order of 1 MeV higher than 
the ones found in the quasi-particle approximation. 
In the latter case we find a saturation point at the edge
of the densities given, in the continuum method this is shifted
to lower densities. 
The reason for this can be found in the next two columns,
$\langle K \rangle/2$ is the average kinetic energy, the 
first contribution to the energy in Eq. \ref{def_energy}.
$\langle V \rangle/2$ is the first moment of the spectral 
function (also known as the mean removal energy), 
the second term in Eq. \ref{def_energy}.
In the continuum method the average kinetic energy is 
larger than in the quasi-particle method. 
This is obvious, in calculating $\langle K \rangle$ we 
integrate over $n(\bar{p})$ with an additional weight $p^2$
over all momenta.
Since strength is moved to higher momenta which have larger 
weights and $\langle K \rangle$ is larger than found
in the quasi-particle approximation.
Moreover, the amount of particles with a momentum larger than 
$p_f$ ($n^+(\bar{p})$) slightly decreases with increasing density.
Altogether this leads to an almost constant difference between
the values of $\langle K \rangle$ calculated in the continuum
and quasi-particle approximation.
For the correlation contribution $\langle V \rangle$ the 
difference larger at 
the lower end of the calculated density range than at the higher end. 
Both effects combine to an effective repulsion which in the 
continuum approach shifts the saturation point to lower densities 
(see also Fig. \ref{fig_binding}).
We have to be careful in generalizing this finding:
the relatively large shift in the saturation point might 
be a result of the system being only slightly bound which makes the
saturation point very sensitive.
Using a realistic interaction the binding will become stronger and 
the saturation point will be less sensitive to small changes in
the self-energy.
On the other hand, the stronger tensor force of such an interaction
leads to a higher depletion and therefore enhances
effects of the continuum contributions.

In the fifth column we give values for the single particle energy
at the Fermi-surface, equal to the chemical potential $\mu$.
In a thermo-dynamically consistent model this is related to 
the binding energy via the Hugenholtz-van Hove theorem \cite{Hugenholtz}:
\begin{equation}
\epsilon_{sp}(p_f) = E_{b} + \frac{P}{\rho},
\label{HvHtheo}
\end{equation}
where $P$ is the pressure of the system, by definition 
$P \equiv \rho^2 {\partial E_{b}}/{\partial \rho}$.
An obvious special case is that at saturation the chemical potential
should be equal to the binding energy. 
The difference between these two is a good measure of the 
thermo-dynamical consistency of the model. 
As we already noted above, the T-matrix approximation is 
a thermo-dynamically consistent approximation \cite{Baym}.
Since self-consistency is a crucial requirement for the 
thermo-dynamical consistency
and the improved degree of self-consistency over
the quasi-particle approximation 
one might expect improvements also with regard to the Hugenholtz-van Hove theorem.
However, this is not the case, we find similar violations in both 
approaches: a violation at saturation of 5 MeV 
for both calculations.
This is somewhat larger than we found for 
a relativistic calculation in the quasi-particle approximation
\cite{FdJ_cons,FdJ_consII}, but in the same range.
Although it is still far from ideal, it is already much better 
than the typical values found in the Brueckner approximation (> 20 MeV). 

In the two rightmost columns we give values for the 
density calculated using Eq. \ref{def_density} divided by the density 
calculated in the 
quasi-particle approximation ($\frac{I}{6\pi^2} p^3_f$), 
divided in contributions with momenta below ($n^-(\bar{p})$)
and momenta above ($n^+(\bar{p})$) the Fermi-momentum. 
The momentum integration was carried out up to 3$p_f$, there
$n(\bar{p})$ is already much smaller than the accuracy with which
we fullfill the spectral function sum-rule Eq. \ref{def_sumrule}.
Particle conservation requires that the sum of the 
two contributions should be equal to unity. 
In our case the particle sum-rule is slightly too
high, by a couple of promilles for both calculations
with no significant difference between the two. 
Although this deviation is within our numerical accuracy there seems to be a 
trend that at higher Fermi-momenta, the calculated relative density is
lower. 
A similar trend is also reported in the calculations of 
Refs \cite{Ramos,VonderfechtII}.

Given the crucial role of self-consistency in either thermodynamical
consistency and conservation laws we find it surprising 
that we do not find significant differences between the
violations in both the quasi-particle approximation and 
the continuum approach. 
Regarding self-consistency the continuum approach clearly goes
beyond the quasi-particle approximation. 
A fully self-consistent calculation would include a spreading of
the quasi-particle peak. 
Since the widths involved there are not too large (most of the particles are 
around the Fermi-surface where the width is small), we think the effect
of this would be smaller than the continuum contributions we included.
Presuming this, it seems that (the lack of) self-consistency is not the
reason for the violations we find.
We already touched upon one possible reason, the validity of the 
gradient expansion. 
In Ref. \cite{Botermans} the conservation laws are proven in momentum
space only within the gradient expansion,
and one might speculate that the violations we find are caused by higher
order derivatives of the $T$-matrix.
We will investigate this point and possible improvements in the future. 

\section{Conclusions}

We presented a calculation for nuclear matter in which we went beyond
the usual quasi-particle approximation and included off-shell effects
in the self-consistent determination of the self-energies.
We achieved this by a suitable approximation of the spectral function
that preserves important properties like the sum-rule 
and the zero at an energy equal the chemical potential.
For momenta below the Fermi-momentum (holes) our approximation amounted to 
putting all hole-strength into the quasi-particle peak and retaining the 
tail of the spectral function with particle strength. 
For particles a similar procedure was followed.
It resulted in a model 
in which the additional continuum contributions are a perturbation on 
the quasi-particle contributions:
in the limit of zero imaginary part of the
self-energy the continuum contributions vanish and the model reduces 
to the conventional quasi-particle approximation. 

We presented results using a non-relativistic version of the Bonn
potential in which we reduce the tensor force to avoid bound states. 
From the model calculation we found that the quasi-particle spectrum 
changes a few MeV upon incorporating the continuum contributions, the most 
for small momenta. 
The average strength typically present in the tail of the spectral function
we now include is in the order of 6 \%.
In the equation of state we found that the continuum contributions
provide an additional repulsion: the saturation point 
is shifted to a lower density.
We also checked particle conservation and thermo-dynamical consistency
as expressed by the Hugenholtz-van Hove theorem.
Since self-consistency is of prime importance in the proofs of these
one would a-priori expect that the violations of these 
should diminish in the continuum approach because of the increase 
in self-consistency over the quasi-particle approximation.
However, we found very similar violations for both approaches.
We indicated that these might not be caused by lack of self-consistency
but by higher order derivatives of the $T$-matrix, 
which are ignored in the gradient expansion.
A better understanding and possible improvements will be subject of 
future studies. 
Another point of future attention must be the inclusion of bound states in 
the approach. 
This will allow to perform a calculation with a realistic interaction,
these typically generate a depletion of $\sim$ 15 \% and one 
expects the effects of the continuum contributions to be even larger.

\appendix
\section*{Calculation of the self-energies}

All integrations are transformed to 
c.m. defining variables $P$ and $p$.
The incoming momentum is $(\omega_{lab}, p_{lab})$, the momentum
over which is integrated is $(\omega_k, k)$,
so that $\bar{P} = \bar{p}_{lab} + \bar{k}$,
$\bar{p} = (\bar{p}_{lab} - \bar{k})/2$.
The first contribution to the self-energy in Eq. \ref{def_sigma} 
(the Brueckner type term) becomes:
\begin{eqnarray}
\lefteqn{T^{(+)}g^<(\omega_{lab}, p_{lab}) = }
\nonumber\\
& &
\int_{P^h_{min}}^{P^h_{max}} dP^2 \int_{p_{min}}^{p_{max}} dp^2 
\frac{1}{(2\pi)^2 2p_{lab}}
\langle p | T^{(+)}(\omega_{lab} + \omega_{k,qp}) | p \rangle Z_h(k) +
\nonumber\\
& &
\int_{P^p_{min}}^{P^p_{max}} dP^2 \int_{p_{min}}^{p_{max}} dp^2 
\frac{1}{(2\pi)^2 2p_{lab}}
\int^\mu_{-\infty} \frac{d \omega_k}{2 \pi}
\langle p | T^{(+)}(\omega_{lab} + \omega_k) | p \rangle a(\omega, k).
\label{sigma_br}
\end{eqnarray}
The first integral integrates over the momenta $k < p_f$, there the 
spectral function has a quasi-particle pole for $\omega_k = \omega_{k,qp}$.
This is reflected in the limits of the integration:
$P^h_{min} = \max(p_{lab} - p_f, 0), P^h_{max} = p_{lab} + p_f$ and
$p^2_{min,max} = \frac{1}{2}(k^2_{min,max} + p^2_{lab}) - \frac{P^2}{4}$
with $k_{min} = \max(P - p_{lab}, 0), k_{max} = \min(p_f, P + p_{lab})$.
The second integral integrates over the momenta $k > p_f$,
there we have to integrate the energy variable over the 
tail of the spectral function.
Here the integration limits are $P^p_{min} = \max(p_f - p_{lab}, 0),
P^p_{max} = \infty$, the expressions for $p^2_{min,max}$ are the same
as before but now $k_{min} = \max(P - p_{lab}, 0), p_f), k_{max} = \infty$.
In the practical calculation we limit this integration by 
$k_{max} = 2p_f$ which implies $P_{max} = p_{lab} + k_{max}$.

In the calculation of $Im[ T^< g^{(-)}]$ we can identify various 
contributions. 
First we integrate separately over hole and particle external
momenta. 
Each of these integrations have a quasi-particle part
and a continuum contribution.
Also $T^<$ has a quasi-particle and a continuum part. 
This all leads to eight different contributions to the
self-energy. 
Note also that all integrations are restricted by the 
Pauli-blocking operator $Q_{hh}$ present in $T^<$, 
specifically these are 
$Q_{hh}(P,p) = 0 \forall P> 2p_f$ and 
$Q_{hh}(P,p) = 0 \forall p^2 > p_f^2 - P^2/4$.

The most important contribution is the quasi-particle contribution
from an integration over an external hole momentum with the 
quasi-particle part from $T^<$, we denote this by a 
subscript {\it qp,qp,h}
\begin{equation}
Im[ T^< g^{(-)}(\omega_{lab}, \bar{p}_{lab})]_{qp,qp,h} = 
\int_{P_{min}}^{P_{max}} dP^2 \int_{p_{min}}^{p_{max}} dp^2 
\frac{1}{(2\pi)^2 4p_{lab}}
\langle p | T^<_{qp}(\omega_{lab} + \omega_{k,qp}) | p \rangle Z_h(k).
\label{rea_qpqph}
\end{equation}
The integration limits are now
$P_{min} = \max(p_{lab} - p_f, 0), P_{max} = \min(p_{lab} + p_f, 2p_f)$,
$p_{min,max}$ are the same as above and
$k_{min} = \max(P - p_{lab}, 0), k_{max} = \min(p_f, P + p_{lab})$.
Recall that we also have the condition 
$\omega_{lab} + \omega_{k,qp} = 
\frac{P^2}{4m} + \frac{p^2_0}{m} + 2\Sigma(P,p_0)_{qp}$.
Since the values which $p_0$ can assume are limited by $Q_{hh}$
the region where the integrand is finite is further limited. 
We implement this numerically.

We also have the equivalent contribution where we integrate over an
external particle momentum
\begin{equation}
Im[ T^< g^{(-)}(\omega_{lab}, \bar{p}_{lab})]_{qp,qp,p} = 
\int_{P_{min}}^{P_{max}} dP^2 \int_{p_{min}}^{p_{max}} dp^2 
\frac{1}{(2\pi)^2 4p_{lab}}
\langle p | T^<_{qp}(\omega_{lab} + \omega_{k,qp}) | p \rangle Z_p(k),
\label{rea_qpqpp}
\end{equation}
for which we have the integration limits
$P_{min} = \max(p_f - p_{lab}, 0),
P_{max} = 2p_f$ and $k_{min} = \max(P - p_{lab}, 0), p_f), k_{max} = \infty$.
Again in the practical calculation we limit this integration by 
$k_{max} = 2p_f$ which implies $P_{max} = \min(2p_f, p_{lab} + k_{max})$.
The integration region is further limited by the limitations
on $p_0$.
From Eq. \ref{t_small} it is clear that $T^<$ is zero for $\Omega \geq 2\mu$.
This results in the contribution above being zero for $\omega_{lab} = \mu$ since
for particles $\omega_{k,qp} \geq \mu$.
 
A second class are the contributions where the external integration
is over a tail of a spectral function with the quasi-particle 
contribution of $T^<$
\begin{eqnarray}
\lefteqn{
Im[ T^< g^{(-)}(\omega_{lab}, \bar{p}_{lab})]_{qp,cont,h} =}
\nonumber\\
& &
\int_{P_{min}}^{P_{max}} dP^2 \int_{p_{min}}^{p_{max}} dp^2 
\frac{1}{(2\pi)^2 4p_{lab}}
\int_\mu^{2\mu - \omega_{lab}} \frac{d \omega_k}{2 \pi}
\langle p | T^<_{qp}(\omega_{lab} + \omega_k) | p \rangle a(\omega, k),
\label{rea_qpconth}
\end{eqnarray}
and the integration limits are similar as for the 
{\it qp,qp,h} contribution, Eq. \ref{rea_qpqph},
$P_{min} = \max(p_{lab} - p_f, 0), P_{max} = \min(p_{lab} + p_f, 2p_f)$,
$p_{min,max}$ are the same as above and
$k_{min} = \max(P - p_{lab}, 0), k_{max} = \min(p_f, P + p_{lab})$.
From the energy integration limits we see that
this contribution is zero for $\omega_{lab} \geq \mu$. 
The particle momentum counterpart is:
\begin{eqnarray}
\lefteqn{
Im[ T^< g^{(-)}(\omega_{lab}, \bar{p}_{lab})]_{qp,cont,p} =}
\nonumber\\
& &
\int_{P_{min}}^{P_{max}} dP^2 \int_{p_{min}}^{p_{max}} dp^2 
\frac{1}{(2\pi)^2 4p_{lab}}
\int_{-\infty}^{\mu_{max}} \frac{d \omega_k}{2 \pi}
\langle p | T^<_{qp}(\omega_{lab} + \omega_k) | p \rangle a(\omega, k).
\label{rea_qpcontp}
\end{eqnarray}
With the same integration limits as the {\it qp,qp,p} contribution Eq. 
\ref{rea_qpqpp} and $\mu_{max} = \min(2 \mu -  \omega_{lab}, \mu)$.

The third category are the contributions where we have the 
continuum contribution of $T^<$ with an external quasi-particle
part
\begin{equation}
Im[ T^< g^{(-)}(\omega_{lab}, \bar{p}_{lab})]_{cont,qp,h} = 
\int_{P_{min}}^{P_{max}} dP^2 \int_{p_{min}}^{p_{max}} dp^2 
\frac{1}{(2\pi)^2 4p_{lab}}
\langle p | T^<_{cont}(\omega_{lab} + \omega_{k,qp}) | p \rangle Z_h(k).
\label{rea_contqph}
\end{equation}
The integration limits are 
$P_{min} = \max(p_{lab} - p_f, 0), P_{max} = p_{lab} + p_f$,
$p_{min,max}$ are the same as above and
$k_{min} = \max(P - p_{lab}, 0), k_{max} = \min(p_f, P + p_{lab})$.
Its counterpart with a particle momentum quasi-particle contribution is
\begin{equation}
Im[ T^< g^{(-)}(\omega_{lab}, \bar{p}_{lab})]_{cont,qp,h} = 
\int_{P_{min}}^{P_{max}} dP^2 \int_{p_{min}}^{p_{max}} dp^2 
\frac{1}{(2\pi)^2 4p_{lab}}
\langle p | T^<_{cont}(\omega_{lab} + \omega_{k,qp}) | p \rangle Z_p(k).
\label{rea_contqpp}
\end{equation}
In the practical calculation we have the integration limits
$P_{min} = \max(p_f - p_{lab}, 0),
P_{max} = p_{lab} + k_{max})$ and 
$k_{min} = \max(P - p_{lab}, 0), p_f), k_{max} = 2p_f$.
Using the same argument as for the contribution Eq. \ref{rea_qpqpp}
it follows that this contribution is also zero for $\omega_{lab} = \mu$.

In the fourth and final category we have the continuum contributions
from both $T^<$ and the external integration:
\begin{eqnarray}
\lefteqn{
Im[ T^< g^{(-)}(\omega_{lab}, \bar{p}_{lab})]_{cont,cont,h} =}
\nonumber\\
& &
\int_{P_{min}}^{P_{max}} dP^2 \int_{p_{min}}^{p_{max}} dp^2 
\frac{1}{(2\pi)^2 4p_{lab}}
\int_\mu^{2\mu - \omega_{lab}} \frac{d \omega_k}{2 \pi}
\langle p | T^<_{qp}(\omega_{lab} + \omega_k) | p \rangle a(\omega, k).
\label{rea_contconth}
\end{eqnarray}
The integration limits are the same as in Eq. \ref{rea_contqph}.
From the integration limits it again follows that this contribution 
is zero for $\omega_{lab} = \mu$.
The particle-momentum counterpart is:
\begin{eqnarray}
\lefteqn{
Im[ T^< g^{(-)}(\omega_{lab}, \bar{p}_{lab})]_{cont,cont,p} =}
\nonumber\\
& &
\int_{P_{min}}^{P_{max}} dP^2 \int_{p_{min}}^{p_{max}} dp^2 
\frac{1}{(2\pi)^2 4p_{lab}}
\int_{-\infty}^{\mu_{max}} \frac{d \omega_k}{2 \pi}
\langle p | T^<_{qp}(\omega_{lab} + \omega_k) | p \rangle a(\omega, k).
\label{rea_contcontp}
\end{eqnarray}
The integration limits are the same as in Eq. \ref{rea_contqpp} and
$\mu_{max} = \min(2 \mu -  \omega_{lab}, \mu)$.

In Eq. \ref{zero_im_sig} we already presented the important condition
that the imaginary part of the self-energy is zero at
the chemical potential $\mu$. 
In our calculation the imaginary part of the Brueckner type
contribution Eq. \ref{sigma_br} is canceled by the various imaginary 
parts of $T^< g^{(-)}$.
We can detail this further, we already identified the contributions to
$Im[ T^< g^{(-)}]$ which are zero at $\omega_{lab} = \mu$.
Using the unitarity relation of $T^{(+)}$ we can identify four 
contributions to the imaginary part of the Brueckner-type contribution.
The first contribution of Eq. \ref{sigma_br} has a quasi-particle
part integrated over a hole momentum, this is canceled by 
Eq. \ref{rea_qpqph}, the continuum part integrated over a hole
momentum is canceled by Eq. \ref{rea_contqph}.
The integration over a particle line of the quasi-particle contribution
is canceled by Eq. \ref{rea_qpcontp} and the continuum part is 
canceled by Eq. \ref{rea_contcontp}.

\begin{figure}
\caption{The quasi-particle spectrum calculated at normal nuclear
matter density in the two approximations described in the 
text. 
The real and imaginary part of the self-energies at the quasi-particle 
peak in the continuum approach are represented
by the solid and dotted lines respectively. 
The dashed and dash-dotted line stand for the real and imaginary part
calculated in the quasi-particle approximation. 
}
\label{qp_spectrum}
\end{figure}

\begin{figure}
\caption{The real (lower panel) and imaginary (upper panel) of the 
self-consistent self-energy calculated in the continuum approach as a 
function of energy and momentum. 
For reasons of visibility the points of view are chosen differently
and the imaginary part was multiplied by -1.
Note the zero of the imaginary part at the chemical potential and the
structure this induces in the real part. 
}
\label{full_sig}
\end{figure}

\begin{figure}
\caption{The energy dependence of the self-energy for momentum 
equal to the Fermi momentum. 
The solid line is the real part of the self-energy in the continuum
approach, the dashed line the real part of the self-energy in the 
quasi-particle approximation. 
The dotted and dash-dotted line are the imaginary parts of the
self-energy calculated in the continuum and in the quasi-particle
approximation respectively. On the scale of the figure the latter are 
almost identical. 
On the horizontal axis we use the variable $\omega'$, defined as the
distance to the quasi-particle peak.
}
\label{sig_omega}
\end{figure}

\begin{figure}
\caption{The binding energy and its components calculated in
the two approximations discussed in the text.
The solid line is the binding energy in the continuum approximation,
the dashed line the binding energy in the quasi-particle approximation.
The dotted and dash-dotted line stand for the kinetic contributions in the
continuum and quasi-particle approximation respectively.
The correlation contributions in the continuum and quasi-particle 
approximation are represented by the dot-dot-dash and the short dashed lines
respectively.}
\label{fig_binding}
\end{figure}

\begin{table}
\begin{tabular}{ccccccc}
$p_f$ & $E_b$ & $\frac{\langle K \rangle}{2}$ & $\frac{\langle V \rangle}{2}$
      & $\epsilon_{sp}(p_f)$ 
      & $n^-(\bar{p})$ & $n^{+}(\bar{p})$ \\
(GeV/c)& (MeV) & (MeV) & (MeV) & (MeV) & (-) & (-) \\
\hline
.25 & -4.2 & 10.0 & -13.2 & -12.5 & 0.935 & 0.068 \\
.27 & -4.8 & 11.7 & -16.5 & -12.6 & 0.942 & 0.061 \\
.29 & -5.2 & 13.5 & -18.7 & -11.7 & 0.946 & 0.057 \\
.31 & -5.3 & 15.4 & -20.6 & -10.0 & 0.947 & 0.054 \\
\end{tabular}
\caption{Various observables as explained in the text 
calculated in the quasi-particle approximation.}
\label{qp_observables}
\end{table}

\begin{table}
\begin{tabular}{ccccccc}
$p_f$ & $E_b$ & $\frac{\langle K \rangle}{2}$ & $\frac{\langle V \rangle}{2}$
      & $\epsilon_{sp}(p_f)$ 
      & $n^-(\bar{p})$ & $n^{+}(\bar{p})$ \\
(GeV/c)& (MeV) & (MeV) & (MeV) & (MeV) & (-) & (-) \\
\hline
.25 & -3.4 & 11.9 & -15.3 & -10.7 & 0.934 & 0.069 \\
.27 & -3.8 & 13.6 & -17.4 & -10.5 & 0.942 & 0.060 \\
.29 & -4.1 & 15.5 & -19.6 & -\phantom{1}9.5 & 0.946 & 0.056 \\
.31 & -3.9 & 17.4 & -21.3 & -\phantom{1}7.3 & 0.947 & 0.054 \\
\end{tabular}
\caption{Various observables as explained in the text 
calculated in the continuum approximation.}
\label{cont_observables}
\end{table}

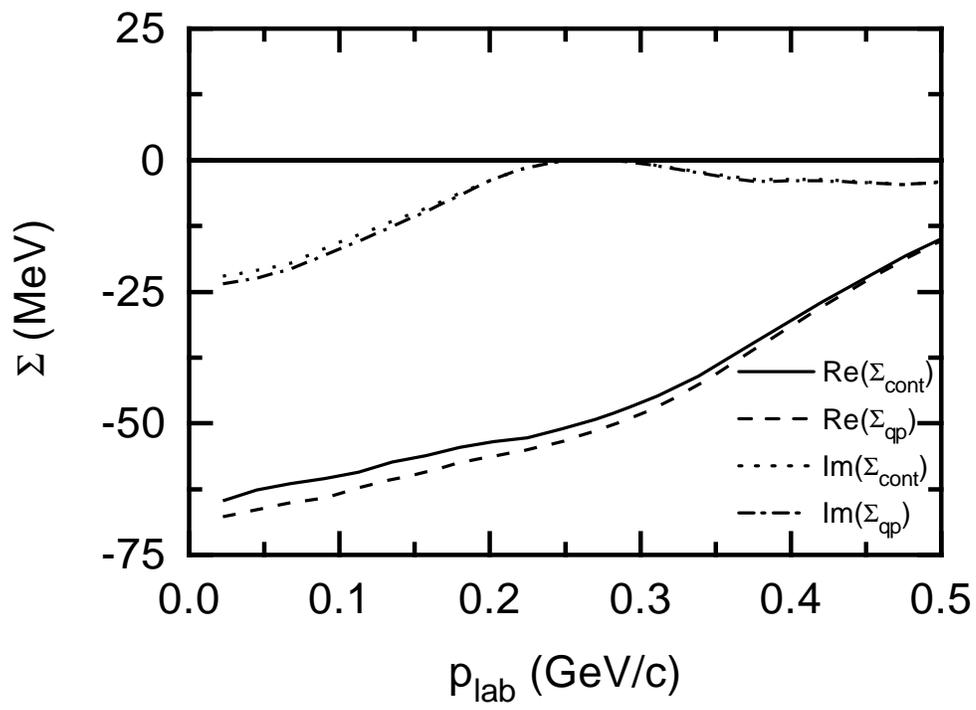
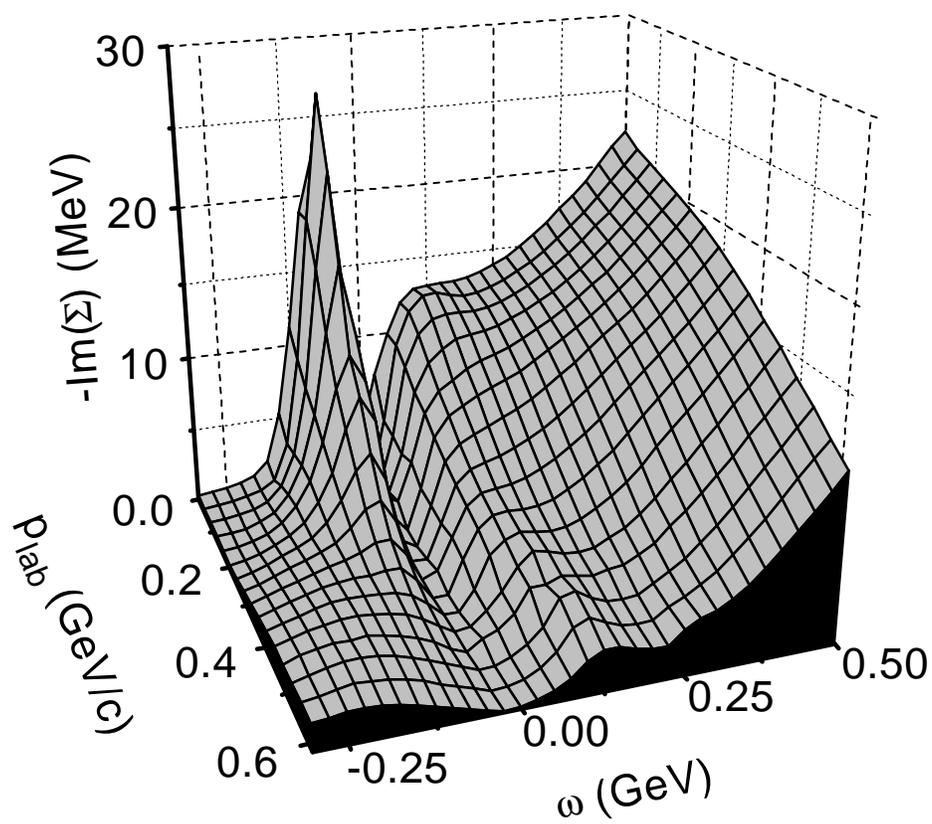
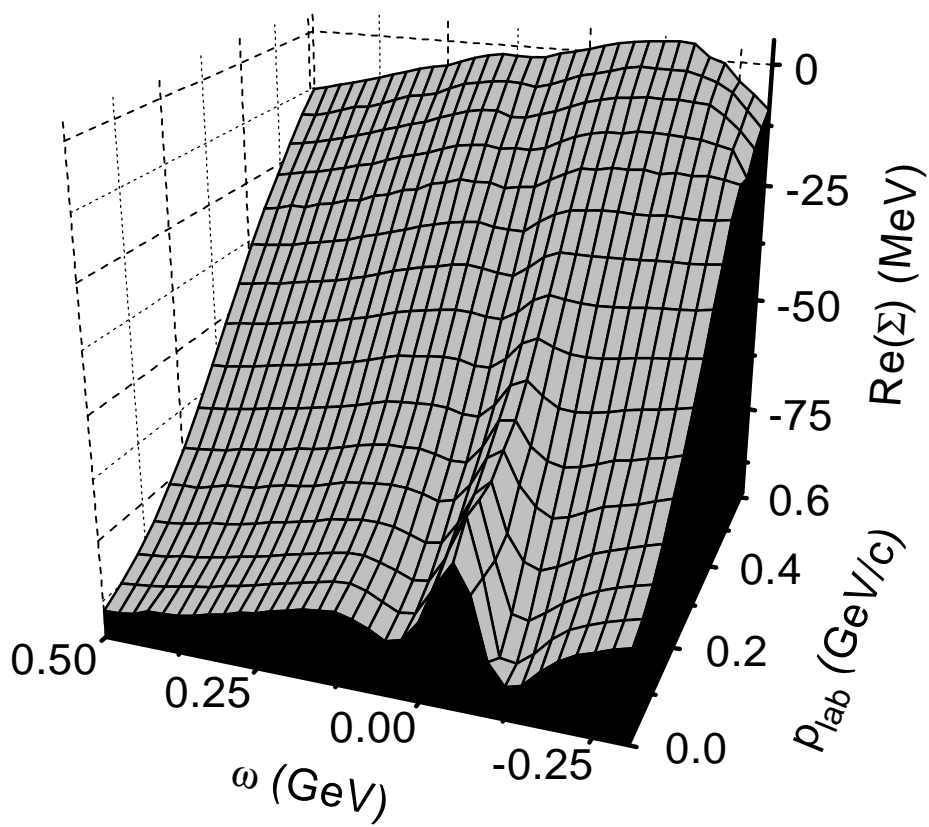
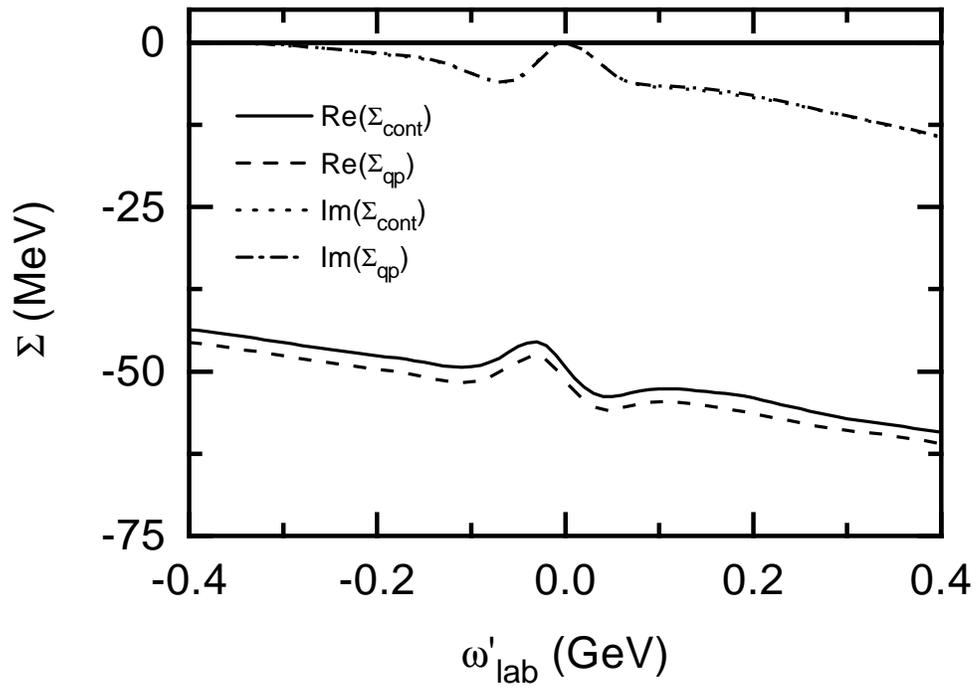
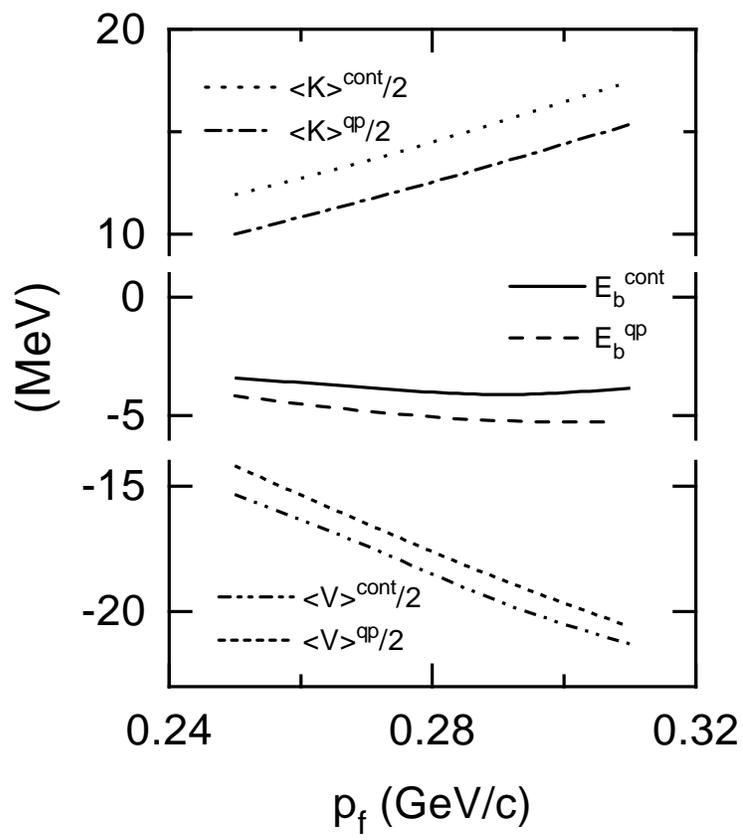
\end{document}